\documentclass[useAMS,usenatbib,amsmath,amssymb]{mn2e}

\usepackage{times}
\usepackage{graphicx}
\usepackage{subfigure}

\newcommand{\beq}{\begin{equation}}
\newcommand{\eeq}{\end{equation}}

\def\gs{\mathrel{\lower0.6ex\hbox{$\buildrel {\textstyle >}\over{\scriptstyle \sim}$}}}
\def\ls{\mathrel{\lower0.6ex\hbox{$\buildrel {\textstyle <}\over{\scriptstyle \sim}$}}}

\newcommand{\aap}{A\&A}
\newcommand{\apj}{ApJ}
\newcommand{\apjl}{ApJ}

\newcommand{\prd}{Phys. Rev. D}
\newcommand{\prl}{Phys. Rev. Lett.}
\newcommand{\mnras}{MNRAS}

\newcommand{\nat}{Nature}

\begin{document}

\title[Cosmography with lensed LISA GW sources]{Cosmography with strong lensing of LISA gravitational wave sources}
\author[Sereno et al.]{
M. Sereno$^{1,2}$\thanks{E-mail: mauro.sereno@polito.it (MS)}, Ph. Jetzer$^{3}$, A. Sesana$^{4}$, M. Volonteri$^{5}$
\\
$^1$Dipartimento di Fisica, Politecnico di Torino, Corso Duca degli Abruzzi 24, 10129 Torino, Italia\\
$^2$INFN, Sezione di Torino, Via Pietro Giuria 1, 10125, Torino, Italia\\
$^{3}$Institut f\"{u}r Theoretische Physik, Universit\"{a}t Z\"{u}rich, Winterthurerstrasse 190, 8057 Z\"{u}rich, Switzerland\\
$^{4}$Max Planck Institute for Gravitationalphysik (Albert Einstein Institute), Am M\"uhlenberg, 14476, Golm, Germany\\
$^{5}$Department of Astronomy, University of Michigan, Ann Arbor, MI 48109, USA
}

%\date{April 4, 2011}

\maketitle

\begin{abstract}
LISA might detect gravitational waves from mergers of massive black hole binaries strongly lensed by intervening galaxies \citep{ser+al10}. The detection of multiple gravitational lensing events would provide a new tool for cosmography. Constraints on cosmological parameters could be placed by exploiting either lensing statistics of strongly lensed sources or time delay measurements of lensed gravitational wave signals. These lensing methods do not need the measurement of the redshifts of the sources and the identification of their electromagnetic counterparts. They would extend cosmological probes to redshift $z \ls 10$ and are then complementary to other lower or higher redshift tests, such as type Ia supernovae or cosmic microwave background. The accuracy of lensing tests strongly depends on the formation history of the merging binaries, and the related number of total detectable multiple images. Lensing amplification might also help to find the host galaxies. Any measurement of the source redshifts would allow to exploit the distance-redshift test in combination with lensing methods. Time-delay analyses might measure the Hubble parameter $H_0$ with accuracy of $\gs 10~\mathrm{km~s^{-1}Mpc^{-1}}$. With prior knowledge of $H_0$, lensing statistics and time delays might constrain the dark matter density ($\delta \Omega_\mathrm{M} \gs 0.08$, due to parameter degeneracy). Inclusion of our methods with other available orthogonal techniques might significantly reduce the uncertainty contours for $\Omega_\mathrm{M}$ and the dark energy equation of state.
\end{abstract}


\section{Introduction}

Observation of gravitational waves (GWs) by extragalactic sources is going to open a new window for astronomy. The space-based Laser Interferometer Space Antenna \citep[ LISA]{dan+al96} is expected to observe up to several hundreds of events per year \citep{s05,ses+al07,ses+al10}. The loudest signals at LISA frequencies, $f\sim \mathrm{mHz}$, should originate from coalescing massive black hole binaries (MBHBs) with total masses in the range $10^3$-$10^7~M_\odot$ out to $z\sim 10$-$15$ \citep{hug02,kle+al09}.

Whenever a new experimental set-up to observe the universe starts working, new possibilities open out. In a previous paper, we discussed a potential new channel for LISA science: multiple imaging of GW sources by intervening strong lensing galaxies \citep{ser+al10}. Lensing of distant sources has been long considered as a test for cosmological theories \citep[ and references therein]{ref66,pa+go81,tur90,fuk+al92,koc93,ser02,cha03,se+lo04,ser07,gi+na09,jul+al10}. GW sources might allow for a variation of these classical tests. 

The main novelty of making cosmography with LISA relies on the property of MBHBs of being standard sirens \citep{sch86}. The luminosity distance to the inspiral GWs can be determined with good accuracy and several methods have already been proposed to exploit this property \citep{ho+hu05,bro+al10,sha+al10,hil+al10}. The main idea on the table is to build-up the Hubble diagram. The relation between distance and redshift changes for different cosmological theories or different cosmological parameters. However, the redshift cannot be measured from the analysis of gravitational waves alone. The use of MBHBs as cosmological probes should rely on the identification of the electromagnetic counterpart in order to measure the redshift of the source.

Lensing offers an alternative tool. In the classical Hubble diagram, we compare observed distances with theoretical expectations depending on cosmology and redshift, which has to be measured in an independent way. When performing cosmography with lensing, we compare some quantity inferred from the lensing analysis, which is in general a combination of cosmological distances, with the theoretical expectation, which depends on the cosmological model. To perform these tests we need either the measured distance to the source or the redshift. In usual lensing studies of quasars or radio-sources, we exploit the measured redshifts of the sources \citep{koc93,cha03}. In lensing studies of LISA sources, we should determine the distances to the sources. The identification of the electromagnetic counter-part of the signal is then not necessary to perform the cosmological tests and we can bypass the weakest link in the already proposed cosmographic methods with LISA.

The chances of observing multiple images of the same GW source with LISA are sizeable for a broad variety of formation histories, ranging from $\ls 40\%$ to $\ls 100\%$ for a 5 years mission according to the redshift distribution of the sources and their intrinsic signal to noise ratio \citep{ser+al10}. It is then worthy to investigate what kind of cosmographic tests might be possible with LISA lensing. Strong lensing by ground-based GW detectors was discussed in \citet{wan+al96}, that considered the constraints on the amount of matter density in compact lenses as derived from lensing statistics in the context of advanced LIGO type detectors. Gravitational lensing of GWs was also considered to measure the relative transverse velocity of a source-lens-observer system \citep{ito+al09} or to obtain information about the typical mass of lens objects \citep{yoo+al07} or as a systematic effect hampering the detection of a very weak primordial GW signal \citep{set09}. 

Here, we focus on the LISA mission and its potential to probe cosmological parameters by strong lensing methods. In Sec.~\ref{stat}, we review some basics of lensing of GWs by galaxies. In Sec.~\ref{tran}, we develop a formalism for the computation of probabilities to detect transient lensing events. In Sec.~\ref{dist}, we discuss how to infer the luminosity distance to a multiply imaged GW source. Cosmological tests based on either lensing statistics or time delay measurements of GW sources are introduced in Sec.~\ref{cosm}. Section~\ref{scen} lists the assumptions made for a plausible lensing scenario whereas Secs.~\ref{fore} and ~\ref{know} contain our forecasting of the lensing test accuracy. Section~\ref{conc} is devoted to some final considerations. 

As reference model we consider a flat $\Lambda$CDM model with $\Omega_\mathrm{M}=0.3$, $\Omega_\Lambda=0.7$, and $h=0.7$ where $h$ is the Hubble constant $H_0$ in units of 100~km~s$^{-1}$~Mpc$^{-1}$.

\section{Basics of gravitational lensing}
\label{stat}

The statistics of gravitational lenses have become a standard tool for cosmology \citep[ and references therein]{koc96,cha03,ser05, zh+se08}. The differential probability of a background source to be lensed by a foreground galaxy with velocity dispersion between $\sigma$ and  $\sigma + d\sigma$ in the redshift interval from $z_\mathrm{d}$ to $z_\mathrm{d}+ d z_\mathrm{d}$ is
\beq
\label{stat1}
\frac{d^2 \tau}{d z_\mathrm{d} d \sigma} = \frac{d n}{d \sigma}(z_\mathrm{d},\sigma) s_\mathrm{cr}(z_\mathrm{d},\sigma) \frac{cd t}{d z_\mathrm{d}}(z_\mathrm{d}) ,
\eeq
where $s_\mathrm{cr}$ is the cross section of the deflector and $d n/d \sigma$ is the differential number density of the lens population.

\subsection{Lens mass density}

Departures from spherical symmetry and details of the radial mass distribution of the lens galaxy induce a relatively small effect on lens statistics and are unimportant in altering the cosmological limits \citep{ma+ri93,koc96,mit+al05}. We can therefore approximate early-type galaxies as singular isothermal spheres (SISs). Two images of a compact source aligned with the lens form at $x_{\pm}=y\pm1$ if $y< 1$, with flux magnification $\mu_{\pm}=(1/y)\pm1$. Here, $x$ and $y$ are the image and the source angular position normalised to the angular Einstein radius, $\theta_\mathrm{E}= 4\pi (\sigma/c)^2 D_\mathrm{ds}/{D_\mathrm{s}}$. $D_\mathrm{d}$, $D_\mathrm{ds}$ and $D_\mathrm{s}$ are the angular diameter distances between the observer and the deflector, the deflector and the source and the observer and the source, respectively. 

In GW lensing,  at variance with the usual lensing, we directly observe the wave-form, which is amplified by $A_{\pm}=\sqrt{\mu_{\pm}}$. The delay between the arrival time of the images, $\Delta t=t_- - t_+$, is
\begin{eqnarray}
\label{deltat1}
\Delta t & = & \Delta t_z y , \Delta t_z  \equiv \frac{32 \pi^2}{c} \left( \frac{\sigma}{c}\right)^4 \frac{D_\mathrm{d} D_\mathrm{ds}}{D_\mathrm{s}} (1+z_\mathrm{d})  \\
& \simeq & 15~\mathrm{days}  \left( \frac{\sigma}{200 \mathrm{km/s}}\right)^4 \frac{D_\mathrm{d} D_\mathrm{ds}}{D_\mathrm{s}\times 0.2\mathrm{Gpc}}(1+z_\mathrm{d})y . \nonumber
\end{eqnarray}

\subsection{Lens population}

The lens distribution can be modelled by a modified Schechter function of the form \citep{she+al03}
\beq
\label{stat2}
\frac{d n}{d \sigma}=n_* \left( \frac{\sigma}{\sigma_*}\right)^\alpha \exp \left[ -\left( \frac{\sigma}{\sigma_*}\right)^\beta\right] \frac{\beta}{\Gamma [\alpha/\beta]} \frac{1}{\sigma},
\eeq
where $\alpha$ is the faint-end slope, $\beta$ the high-velocity cut-off and  $n_*$ and $\sigma_*$ are the characteristic number density and velocity dispersion, respectively. An evolving galaxy density can be parameterised with $n_*(z)=n_{*,0}(1+z)^{3-\nu_{n^*}}$ and $\sigma_*=\sigma_{*,0}(1+z)^{\nu_{\sigma^*}}$ \citep{cha07}. For a constant comoving number density, $\nu_{n^*}=\nu_{\sigma^*}=0$.

\subsection{Detection thresholds and bias}

Lens discovery rates are affected by the ability to observe multiple images \citep{koc93,ser+al10}. For optically luminous quasars or radio sources, lensed systems are detected by selecting resolved multiply images; intrinsic variations show up with a time delay in each image. On the other hand, due to the low angular resolution and the transient nature of the sources, lensed GWs in the LISA context are detected as repeated events in nearly the same sky position. 

The source position is limited to an allowed range, $y_\mathrm{min}\le y \le y_\mathrm{max}$, for which multiple images are detectable. The upper limit $y_\mathrm{max}$ depends on the lens mass, the arrival time, the threshold signal to noise ratio ($\mathrm{SNR_\mathrm{th}}$) and the unlensed amplitude of the source ($\mathrm{SNR_{int}}$) \citep{ser+al10}. 

We require that lensing amplification pushes the signal of the second image above threshold, $A_{-}> \mathrm{SNR_\mathrm{th}^{-}}/\mathrm{SNR_{int}}$, which limits the source position to
\beq
y \le y_\mathrm{max} =   \left[  \left( \mathrm{SNR_\mathrm{th}^{-}}/\mathrm{SNR_{int}}\right)^2   + 1 \right]^{-1} .
\eeq 
The minimum $y_\mathrm{min}$ excises the region near the central caustic where wave optics is effective and the interference pattern covers the multiple images. For the LISA wave-band, geometric optics is adequate and we can put $y_\mathrm{min}=0$ \citep{ser+al10}. 

For our forecasting we deal with known properties of the source population so that we do not need to correct for the magnification bias \citep{fu+tu91,fuk+al92,ser+al10}.

\section{Statistics for transient phenomena}
\label{tran}

Due to the finite duration of the survey, lensing statistics of transient phenomena involve accounting for some missing events due to time delay \citep{ogu+al03}. The cross section of a SIS for a time-limited survey is
\beq
\label{tran1}
s_\mathrm{cr}(z_\mathrm{d}, \sigma)= \pi  R_\mathrm{E}^2 \int^{y_\mathrm{max}}_{y_\mathrm{min}}f(\Delta t (y; \sigma, z_\mathrm{d}))y dy,
\eeq
where $R_\mathrm{E}= D_\mathrm{d} \theta_\mathrm{E}$ is the Einstein radius and $f(\Delta t)$ is the fraction of lenses with time delay $\Delta t$ that can be observed. For lasting images, as for quasars or radio-sources, $f(\Delta t) = 1$, 
\beq
s_{\mathrm{cr}}^*=2 \pi  R_\mathrm{E}^2 \left( y_\mathrm{max}^2 -y_\mathrm{min}^2\right).
\eeq

We consider a uniform distribution of arrival times during the time survey, $T_\mathrm{sur}$. If the monitoring is continuous
\beq
\label{tran2}
f (\Delta t) = 1-\frac{\Delta t}{T_\mathrm{sur}}
\eeq
for $\Delta t<T_\mathrm{sur}$ and 0 otherwise \citep{ogu+al03}. The resulting cross section weighted for the arrival time distribution is then
\beq
\label{stat3}
s_\mathrm{cr}= \pi  D_\mathrm{d}^2 \theta_\mathrm{E}^2 \left[ \left( y_\mathrm{max}^2 -y_\mathrm{min}^2 \right) -\frac{2}{3} \frac{\Delta t_z}{T_\mathrm{sur}} \left( y_\mathrm{max}^3 -y_\mathrm{min}^3 \right) \right].
\eeq
In the following, we specialise to the case of geometric optics ($y_\mathrm{min}=0$).

The differential probability $d \tau/d z_\mathrm{d}$ for a source to be lensed by a deflector in the interval between $z_\mathrm{d}$ and $z_\mathrm{d}+dz_\mathrm{d}$ can be obtained integrating the differential probability in Eq.~(\ref{stat1}). For each lens redshift $z_\mathrm{d}$ there is a maximum velocity  dispersion $\sigma_\mathrm{max}$ such that $s_\mathrm{cr}(\sigma)<0$ for $\sigma > \sigma_\mathrm{max}$. In practice, $\sigma_\mathrm{max}$ is quite large ($\gg \sigma_*$) and the corresponding galaxy density is almost null, $dn/d\sigma (\sigma_\mathrm{max}) \sim 0$. When integrating the differential optical depth, we can then take $\sigma_\mathrm{max}\rightarrow \infty$.  We get
\begin{eqnarray}
\frac{d \tau}{d z_\mathrm{d}}& \simeq & n_{*,0} s_{\mathrm{cr}}^* (\sigma_{*,0};z_\mathrm{d}) \frac{c}{H(z_\mathrm{d})}(1+z_\mathrm{d})^{2+\nu_{n_*}+4\nu_{\sigma_*}} 
\frac{\Gamma \left[ \frac{4+\alpha}{\beta} \right]}{\Gamma \left[ \frac{\alpha}{\beta} \right]} \nonumber \\
& \times & \left\{
1
-\frac{2}{3} \frac{\Delta t (\sigma_{*,0},y_\mathrm{max})}{T_\mathrm{sur}}(1+z_\mathrm{d})^{4\nu_{\sigma_*}} \frac{\Gamma \left[ \frac{8+\alpha}{\beta} \right]}{\Gamma \left[ \frac{4+\alpha}{\beta} \right]}
\right\} , \label{tran3}
\end{eqnarray}
where $\Gamma$ is the Euler gamma function.

The total optical depth for multiple imaging of a compact source, $\tau$, gets a quite compact form for a constant comoving density,  $\nu_{n^*}=\nu_{\sigma^*}=0$. After integration,
\begin{equation}
\label{tran4}
\tau  = \frac{F_*}{30} \left[ D_\mathrm{s} (1+z_\mathrm{s})\right]^3 y_\mathrm{max}^2 \left[ 1-\frac{1}{7} \frac{\Gamma \left[ (8+\alpha)/\beta \right]}{\Gamma \left[ (4+\alpha)/\beta \right]}\frac{\Delta t_*}{T_\mathrm{sur}}\right],
\end{equation}
where
\begin{eqnarray}
F_*& = & 16 \pi^3 n_{*,0} \left( \frac{\sigma_{*,0}}{c}\right)^4 \frac{\Gamma \left[ (4+\alpha)/\beta \right]}{\Gamma \left[ \alpha/\beta \right]}  ;\label{tran5}\\
\Delta t_* &= &   32 \pi^2 \left( \frac{\sigma_{*,0}}{c}\right)^4 \frac{D_\mathrm{s}}{c}(1+z_\mathrm{s}) y_\mathrm{max} . \label{tran6}
\end{eqnarray}
The optical depth $\tau$ is a function of the source redshift $z_\mathrm{s}$ and intrinsic SNR (through $y_\mathrm{max}$); the cosmological parameters enter in the angular diameter distances.

\section{Distance determination}
\label{dist}

In a seminal paper, \citet{sch86} showed that measurements of the amplitude, frequency and frequency derivative of the inspiralling massive binary black holes could yield a precise estimate of their luminosity distance. This would allow us to measure the distance in a novel way, making GW sources potentially powerful standard sirens. 

Each harmonic of the inspiral polarisations is inversely proportional to the distance, $p \propto 1/D_\mathrm{L}$, where $D_\mathrm{L}$ is the luminosity distance. For well modelled systems, LISA will be able to measure the luminosity distance (but not the redshift) to massive BH binaries with 1-10\% accuracy. The measurement precision is largely limited by pointing error and weak lensing distortion \citep{la+hu06,kle+al09}. Gravitational lensing will randomly magnify or demagnify MBHB signal, and thus systematically modify any distance measurement \citep{ho+hu05}.

LISA should detect independently two above threshold signals in the same sky position in order to claim lensing. The two lensed wave-forms are identical, apart from an overall factor connected to lensing amplification. For each image $p_{\pm} \propto A_{\pm}/D_\mathrm{L}$, where the suffix refers to the lensing parity of the image and not to the polarisation. 

The distance to the lensed source can be inferred in the following way. The ratio of the amplifications $A_{-}/A_{+}$ can be measured from $p_-/p_+$, and one can directly infer the source position, $y$,
\beq
y=\frac{1-|A_-/A_+|}{1+|A_-/A_+|} .
\eeq 
Once the source position is known, one can quantify the amplification of each image, $A_{\pm}$ and estimate the luminous distance. As a consistency check of the lensing hypothesis one should find that the value of $y$ does not depend on the frequency as is the case instead for the amplitudes. Since lensing amplification can be expressed in terms of the scaled lens position $y$, knowledge of the lens velocity dispersion is not needed.  
Such methodology can be easily generalised to lenses more complex than the simple SIS.

The method above is effective only with regard to strong lensing by a single deflector plane. Weak lensing is well recognised as a potential noise in the determination of the distance to GW sources. Amplification due to large scale structure cannot be filtered out and will still contribute the main uncertainty in the determination of $D_\mathrm{L}$. There are proposals on how this effect could be, at least partially, corrected. Convergence maps reconstructed using galaxy flexion in addition to shear might help to reduce the lensing-induced distance errors by up to 50\% \citep{sha+al10,hil+al10}. A Gaussian distance error with a standard deviation of 10\% was then added to each luminosity distance to simulate the effect of weak lensing errors. A more realistic error should grow with redshift, but we also tested that results are nearly unaffected by assuming a 20\% error.

Magnification or demagnification of signals affect strong lensing observables too \citep{asa98}. However, the universe is quite homogeneous at the high redshifts ($z\gs 10$) probed by LISA sources \citep{wan99}. Magnification effects are also somewhat washed out after averaging over many independent lines of sights \citep{ser+al02}. Lensing statistics are then not heavily affected by the lensing dispersion in the distance-redshift relation \citep{cov+al05}.

\section{Cosmological tests}
\label{cosm}

The lensing optical depth depends on both source redshift and distance. In standard lensing statistics of radio sources or quasars, the source redshift is known and we can estimate the distance assuming a given cosmological model, i.e., a given set of cosmological parameters. Dealing with lensing statistics of observed GWs, we know the luminosity distance to the source and we have to work the other way: given a set of cosmological parameters, we estimate the source redshift and, on turn, the angular diameter distance. 

The analysis of the GW-form allow us to infer the luminosity distance to the source, whereas lensing quantities are written in terms of angular diameter distances. In the general theory of relativity, luminosity and angular diameter distances are related through the Etherington principle, $D_\mathrm{L}=(1+z)^2D$. Throughout, luminosity distances are denoted by the suffix ``L". It is also useful to remember that in a flat universe $D_\mathrm{ds}=D_\mathrm{s}-D_\mathrm{d}(1+z_\mathrm{d})/(1+z_\mathrm{s})$.

\subsection{Lensing statistics}

The likelihood function for lensing statistics can be written as \citep{koc93,cha03,mit+al05}
\beq
\label{like1}
{\cal L}_\mathrm{stat} (\Omega_i)= \prod_{i=1}^{N_\mathrm{U}}(1-\tau_i)\prod_{j=1}^{N_\mathrm{
L}}p_{l,j},
\eeq
where $\left\{ \Omega_i \right\}$ is the set of cosmological parameters that characterises the model under investigation. Lens statistics do not depend on the Hubble constant \citep{koc93,cha03}. $N_\mathrm{L}$ is the number of multiple-imaged sources and $N_\mathrm{U}$ is the number of unlensed sources. $p_{l,j}$ is either $d \tau/ d z_\mathrm{d}$ for known lens redshift or $\tau$ for unknown $z_\mathrm{d}$. In principle, a test based on lens statistics can then be performed even without lens identification.

For given values of the cosmological parameters, the relation between $z$ and $D_\mathrm{L}$ is unique. Given the measured $D_\mathrm{L}$, we can then compute the angular diameter distances and ${\cal L}_\mathrm{stat}$ for each set of cosmological parameters. 

The main source of statistical uncertainty is by far the Poissonian noise due to the small number of events. For our forecasting about the accuracy of lensing statistics, the error on $D_\mathrm{L}$, which is of order of $\ls 10\%$, is then negligible.

\subsection{Time delay}

It has been long known that time delay measurements can be used to constrain cosmological parameters \citep{ref66,sah+al06}. If we measure $\Delta t$ and constrain the lens properties with either lensing or other follow-up imaging or spectroscopic observations, we can get an estimate of $D_{\Delta t} \equiv D_\mathrm{d}D_\mathrm{ds}/D_\mathrm{s}$ and in turn constrain the cosmological parameters. Differently from lens statistics, we need the lens redshift to carry out cosmography with time-delays. For each measured $\Delta t$, we can write a $\chi^2$ contribution as
\beq
\label{like2}
\chi^2_{\Delta t_i} (\Omega_i; h; z_\mathrm{s})   =  \left( \frac{D_{\Delta t}^\mathrm{Obs}-D_{\Delta t} }{\delta D_{\Delta t}^\mathrm{Obs}} \right)^2 +  \left( \frac{D_\mathrm{L}^\mathrm{Obs}-D_\mathrm{L}}{\delta D_\mathrm{L}^\mathrm{Obs}} \right)^2 .
\eeq
In the above equation, the apex ``Obs" denotes the measured value of the corresponding quantity. $D_{\Delta t}^\mathrm{Obs}$ is measured with the lensing analysis; the observed value of the luminosity distance $D_\mathrm{L}^\mathrm{Obs}$ is obtained from the wave-form investigation. $D_{\Delta t}$ and $D_\mathrm{L}$ are functions of the cosmological parameters and of the source redshift. Due to the uncertainty $\delta D_\mathrm{L}^\mathrm{Obs}$ on $D_\mathrm{L}^\mathrm{Obs}$, the source redshift is not perfectly known and has to be considered as a model parameter. The second term in the right-hand side of Eq.~(\ref{like2}) shapes the information on the luminosity distance obtained from the analysis of the signal as a Gaussian prior. Once data analysis of LISA data is really performed, best estimates of luminosity distances to mergers are going to be determined together with their probability distributions. More realistic priors for the time delay likelihood should be shaped after these computed distributions.

The likelihood built on time delays is
\beq
\label{like3}
{\cal L}_{\Delta t} (\Omega_i; h; z_{\mathrm{s},i}) \propto \exp \{-\frac{1}{2}\sum_j^{N_L} \chi^2_{\Delta t_j} \},
\eeq
where the sum extends on the lensed systems. The unknown source redshifts $\left\{ z_{\mathrm{s},i} \right\}$ are $N_\mathrm{L}$ additional model parameters. The cosmographic approach with time-delays is not affected by Poissonian noise, whereas the error on $D_\mathrm{L}$ is an important source of statistical uncertainty. The main cosmological dependence is on the Hubble constant, since distances $\propto 1/H_0$.

\section{A scenario for lensing}
\label{scen}

\begin{figure}
  \resizebox{\hsize}{!}{\includegraphics{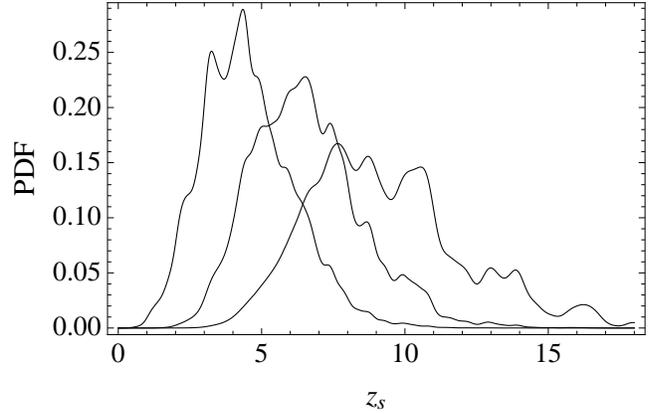}}
   \caption{Probability density functions for the redshift of the 3 lensed sources. PDFs have been obtained by repeatedly extracting a triple from the intrinsic redshift distributions of the simulated LISA sources. Each source was weighted by its optical depth to lensing.}
   \label{fig_pdf_zs_lensed}
\end{figure}

\begin{figure}
  \resizebox{\hsize}{!}{\includegraphics{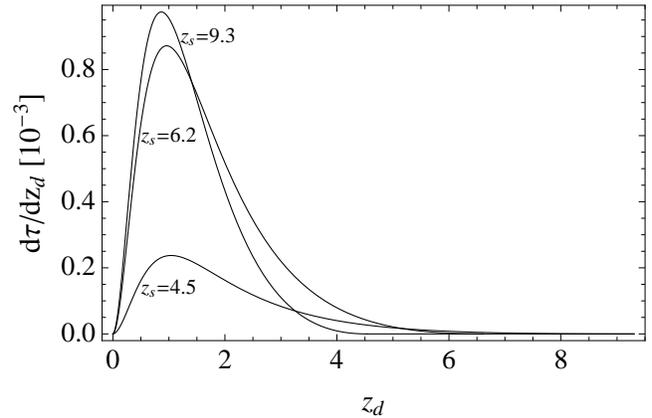}}
 \caption{Differential lensing probability for different source redshifts. Probabilities were obtained through Eq.~(\ref{tran3}). Intrinsic SNR are 400, 12 or 36 for $z_\mathrm{s}=4.5,~6.2$ or $9.3$, respectively.}
   \label{fig_pdf_zd_lenses}
\end{figure}

We tested the predictive power of the lensing methods on a class of flat cosmological models with dark matter, $\Omega_\mathrm{M}$, and a dark energy fluid whose equation of state is parameterised by a constant equation of state $\omega$. Within such context, our ``true" $\Lambda$CDM model is  described by $\Omega_\mathrm{M}=0.3$, and $\omega=-1$. State-of-the art analysis of the cosmic microwave background radiation with informed priors can constrain the Hubble constant with an accuracy $\ls 2\%$ \citep{kom+al10} and estimates should be even more tight by the time LISA is flying. Such error is much smaller than the uncertainties on $D_\mathrm{L}^\mathrm{Obs}$ and $D_{\Delta t}^\mathrm{Obs}$. 

We then consider two frameworks for cosmological parameter forecasting. Either we focus on the determination of the Hubble constant or we take $H_0$ as known and concentrate on what lensing with LISA can do for the dark sector.

\subsection{Deflectors}

A proper modelling of the distribution of the lensing galaxies is central in  lensing statistics. Early-type or late-type populations contribute to the lensing statistics in different ways and type-specific galaxy distributions  are required. As a conservative approach, we did not consider lensing by spiral galaxies. Late-type galaxies contribute no more than 20-30\% of the total lensing optical depth and the knowledge of their number density is plagued by large uncertainties \citep{cha03,mit+al05}. 

In our analysis we used the results of Choi et al. (2007) who analysed data  from the SDSS Data Release 5 to derive the velocity dispersion distribution function of early-type galaxies. They modelled the galaxy population as a modified Schechter function with $n_{*,0} = 8.0{\times} 10^{-3} h^3$~Mpc$^{-3}$, $\sigma_{*,0}=144 {\pm} 5$~km~s$^{-1}$, $\alpha=2.49 \pm 0.10$, and $\beta = 2.29 \pm 0.07$. We keep constant the comoving number density of galaxies.

Even if the LISA angular resolution is quite poor, $\sim 30'$-$1\deg$, lenses are expected to be very massive and luminous. In case the deflector can be identified and its redshift measured, the corresponding Einstein radius can be estimated and one can accurately determine the angular position of the source. This way one could identify the source (more precisely its images) as well with follow-up observations and thus get also its redshift.

\subsection{Sources and lenses}

The number of detectable multiple images depends on the build-up formation history \citep{ser+al10}. Massive mergers at high redshifts can produce very loud GW emission with a noticeable optical depth to lensing but are expected to be quite rare. On the other hand, minor mergers might be more frequent but with a lesser lensing probability due to intermediate redshift and lower intrinsic signal-to-noise. The total lensing probability balances the total number of events, from few dozens to several hundreds detectable coalescences per year \citep{ses+al10}, and their intrinsic loudness. Throughout, the monitoring time for LISA is fixed to $T_\mathrm{sur}=5~\mathrm{years}$. The expected number of lensing events goes from $\ls 1$ to $\ls 4$ \citep{ser+al10}.

We considered two scenarios for the abundance of LISA sources. In a first pessimistic case, we derived cosmological constraints from a single lensing detection. In a more optimistic scenario,  three multiple image events might be seen by LISA. This is the case for example for some hybrid formation histories, such as the ``HybridII" model of \citet{ser+al10}. Mergers were generated in the reference $\Lambda$CDM model. For each source we computed the lensing probability. As a threshold for detection we took $\mathrm{SNR_{th}}=8$. A mean of $\sim2.8$ detectable lensing events is expected. 

We extracted the properties of the three lensed sources from their parent distribution, weighting each source by its optical depth to lensing. The probability density function for the source redshifts (sorted according to their $z$) is plotted in Fig.~\ref{fig_pdf_zs_lensed}. Accordingly to the mean properties of the distributions, we fixed the three source redshifts for our forecasting at $z_\mathrm{s} \simeq 4.5$, $6.2$ and $9.3$.

The corresponding lens redshifts were extracted as the $z_\mathrm{d}$ which maximises $d\tau/dz_\mathrm{d}$ for a given $z_\mathrm{s}$, see Fig.~\ref{fig_pdf_zd_lenses}. We got $z_\mathrm{d} \simeq 0.9$, $1.0$ and $1.0$. Since the shrinking of the cosmological volumes with redshifts, the peak of the probability is always at $z_\mathrm{d} \sim 1$, with no regard to the source redshift. Together with the expectation of being very massive, their moderate redshift might make the lenses identifiable.

For the pessimistic scenario, we computed the most likely lensing configuration in the case of only one lensing event detected in the ``HybridII" formation history. The source is at $z_\mathrm{s} \simeq 7.4$ and the corresponding deflector is at $z_\mathrm{d} \simeq 1.0$.

\section{Forecasting for unknown source redshifts}
\label{fore}

For our forecasting, we exploit different priors according to whether we focus on either $H_0$ or $\Omega_\mathrm{M}$ and $\omega$. When we deal with the Hubble constant, the accelerated expansion of the Universe is assumed to be propelled by a cosmological constant, so that $\omega=-1$ is our delta prior for the dark energy. As far as the dark matter is concerned, as a first case we assume no previous information, i.e., a prior in the form of a uniform distribution, $P(\Omega_\mathrm{M})=1$ for $0 \le \Omega_\mathrm{M} \le1$. Alternatively, we shape results from other methods as a very mild Gaussian prior centred on the ``true" value $\Omega_\mathrm{M}=0.3$ and with a quite large dispersion $\sigma_{\Omega_\mathrm{M}}=0.1$. The prior on $h$ is a flat distribution non null between 0 and 2.

When we focus on the dark sector, $H_0$ is kept to its reference value. As prior for the cosmological parameters we consider a uniform distribution in the square $0 \le \Omega_\mathrm{M} \le 1$ and $-2 \le \omega \le 0$.

\subsection{Lensing statistics}

\begin{figure}
\resizebox{\hsize}{!}{\includegraphics{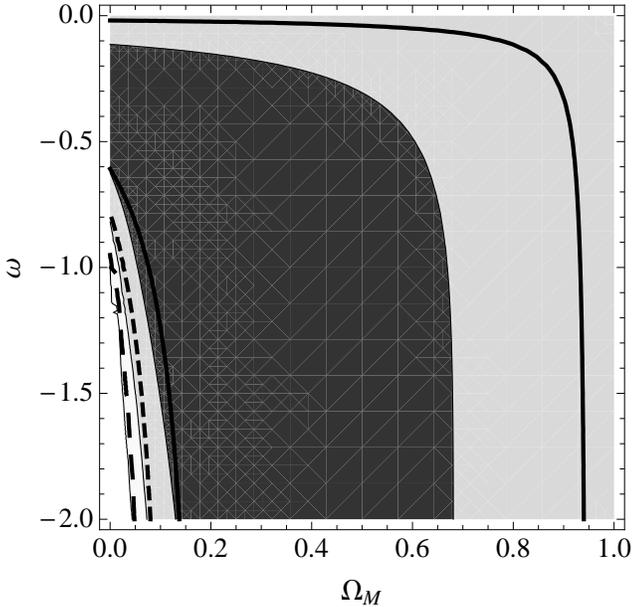}}
 \caption{Posterior probability for $\Omega_\mathrm{M}$ and $\omega$ as derived from lensing statistics in the case of 3 lensed sources. Grey-shadowed regions or thick contours (full, dashed or long-dashed lines) refer to lensing statistics either exploiting information on lens redshift or assuming $z_\mathrm{d}$ to be unknown. Contours are plotted at fraction values $\exp (-2.3/2)$, $\exp(-6.17/2)$, and $\exp(-11.8/2)$ of the maximum, which would denote confidence limit region of 1-, 2- and 3-$\sigma$ in a maximum likelihood investigation, respectively. }
   \label{fig_like_statistics_Omega_M_omega}
\end{figure}

Lensing statistics of GW sources might constrain the dark matter density and the dark energy equation of state. It does not depend on $H_0$. We simulated the measurements of the luminosity distances by adding a Gaussian noise of $10\%$ to the true value. However, since the Poissonian error is the main one, even by doubling the error on the distances the inferred posterior probability of $\Omega_\mathrm{M}$ and $\omega$ from lensing statistics is unaffected in any sensible way. 

The likelihood was computed in a grid. The posterior probability function for the cosmological parameters is plotted in Fig.~\ref{fig_like_statistics_Omega_M_omega}. Lensing statistics can be performed considering either known or unknown lens redshifts. Precise information on $z_\mathrm{d}$ makes the upper limit on the estimated $\Omega_\mathrm{M}$ much stronger. The accuracy on $\Omega_\mathrm{M}$ is $\sim 0.24$ for 3 lenses with measured redshift.

\subsection{Time-delays}

The distance combination $D_{\Delta t}$ is the basic source of information on cosmology from time-delay measurements. The uncertainty in the measurement of $D_{\Delta t}$ depends on the accuracy with which we can measure the time-delay and the lens mass distribution. For transient events, time delay can be measured with great accuracy \citep{ogu+al03} so that the main source of error is from the uncertainties in the mass distribution of the deflector. Lens are expected to be at redshifts $z_\mathrm{d} \sim 1$ and to be very luminous so that follow-up spectroscopy and photometry should be performable. Further constraints come from the lensing modelling exploiting image positions and time-delay. An uncertainty on $D_{\Delta t}$ of order of $20\%$ is a conservative assumption; an uncertainty of $10\%$ is more optimistic and more adequate to the case of a deep follow-up.

When considering time-delay, together with the cosmological parameters we have to consider as unknown model parameters the redshifts of the lensed sources. For our optimistic (pessimistic) lensing scenario, we have then to add 3 (1) unknown source redshifts to the cosmological parameters to characterise the model.

\subsubsection{$H_0$}

\begin{table}
\begin{tabular}[c]{ccccc}       
\hline        
\noalign{\smallskip}
lenses	&$\delta D_{\Delta t}/ D_{\Delta t}$	& $z_\mathrm{s}$	&$P(\Omega_\mathrm{M})$	&$\delta h$ \\
\noalign{\smallskip}
\hline              
\noalign{\smallskip}
3	&10\%	&known		&$\Omega_\mathrm{M}=0.3\pm 0.1$	&$0.11$	\\ 
3	&10\%	&known		&$0\le \Omega_\mathrm{M}\le 1$		&$0.21$	\\ 
3	&10\%	&unknown	&$\Omega_\mathrm{M}=0.3\pm 0.1$	&$0.12$	\\ 
3	&10\%	&unknown	&$0\le \Omega_\mathrm{M}\le 1$		&$0.23$	\\ 
\hline
3	&20\%	&known		&$\Omega_\mathrm{M}=0.3\pm 0.1$	&$0.11$	\\ 
3	&20\%	&known		&$0\le \Omega_\mathrm{M}\le 1$		&$0.24$	\\ 
3	&20\%	&unknown	&$\Omega_\mathrm{M}=0.3\pm 0.1$	&$0.20$	\\ 
3	&20\%	&unknown	&$0\le \Omega_\mathrm{M}\le 1$		&$0.26$	\\ 
\noalign{\smallskip}
\hline              
\noalign{\smallskip}
1	&10\%	&known		&$\Omega_\mathrm{M}=0.3\pm 0.1$	&$0.13$	\\ 
1	&10\%	&known		&$0\le \Omega_\mathrm{M}\le 1$		&$0.30$	\\ 
1	&10\%	&unknown	&$\Omega_\mathrm{M}=0.3\pm 0.1$	&$0.18$	\\ 
1	&10\%	&unknown	&$0\le \Omega_\mathrm{M}\le 1$		&$0.31$	\\ 
\hline
1	&20\%	&known		&$\Omega_\mathrm{M}=0.3\pm 0.1$	&$0.14$	\\ 
1	&20\%	&known		&$0\le \Omega_\mathrm{M}\le 1$		&$0.31$	\\ 
1	&20\%	&unknown	&$\Omega_\mathrm{M}=0.3\pm 0.1$	&$0.34$	\\ 
1	&20\%	&unknown	&$0\le \Omega_\mathrm{M}\le 1$		&$0.38$	\\ 
\hline
\end{tabular}       
\centering       
\caption{Predicted uncertainty on the Hubble constant (column~5) under different hypotheses for number of observed lenses (first column), uncertainty on $ D_{\Delta t}$ (second column), previous knowledge of the lens redshifts (col.~3) and prior on $\Omega_\mathrm{M}$ (col.~4; a Gaussian prior centered on 0.3 with dispersion of 0.1 or a uniform distribution between 0 and 1).}        
\par\noindent 
\label{tab_delta_h}     
\end{table}

Forecasting for the Hubble constant is summarised in Table~\ref{tab_delta_h}. The quoted uncertainty on $H_0$ is the standard deviation of the marginalised posterior probability density. For each case, the parameter space ($H_0$ and $\Omega_\mathrm{M}$ plus the source redshifts) was explored running 4 Markov chains of $2.5\times 10^4$ samples each. Chain convergence was checked through the Gelman and Rubin ratio \citep{ge+ru92,le+br02}, which was well under 1.05 for each parameter.

Even a mild prior on $\Omega_\mathrm{M}$ is very helpful in breaking the degeneracy between $H_0$ and the density parameters. With 3 lenses modelled with deep follow-up observations, an accuracy of $\delta h \sim 0.12$ can be achieved. This figure is larger than present uncertainty but it would provide a direct test on $H_0$ without need for calibration or distance scale ladder in an unexplored redshift range, $z\ls 10$. Even a single lens detection could provide interesting limits, $\delta h \sim 0.20$ or $0.26$ according to the a-priori degree of knowledge on the dark matter density.

\subsubsection{Dark energy} 

\begin{figure}
  \resizebox{\hsize}{!}{\includegraphics{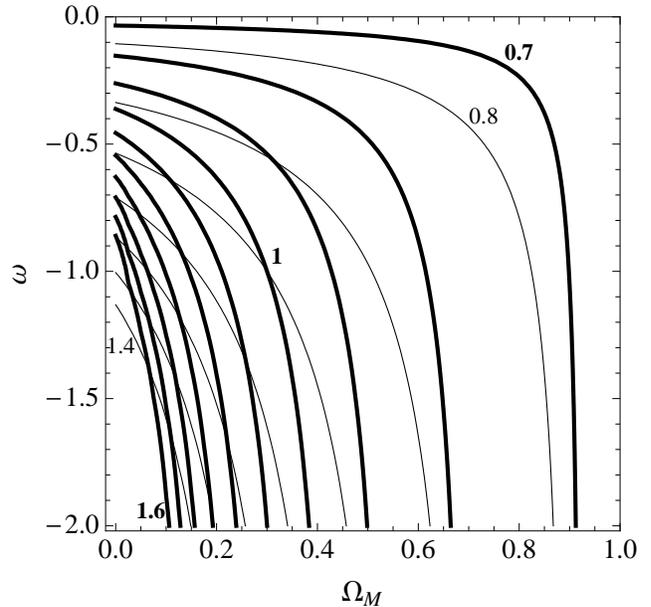}}
   \caption{$D_{\Delta t} (z_\mathrm{d}=1,z_\mathrm{s}=7)$ (thick lines) and $D_\mathrm{s}(z_\mathrm{s}=1)$ (thin lines) as a function of the cosmological parameters. Values of $D_{\Delta t}$ ($D_\mathrm{s}$) are normalised to their value at $\Omega_\mathrm{M}=0.3$ and $\omega=-1$ and run from 0.7 (0.8) to 1.6 (1.4) in steps of 0.1. The Hubble constant is fixed to $h=0.7$.}
   \label{fig_distances_OmegaM_omega}
\end{figure}

\begin{figure}
 \resizebox{\hsize}{!}{\includegraphics{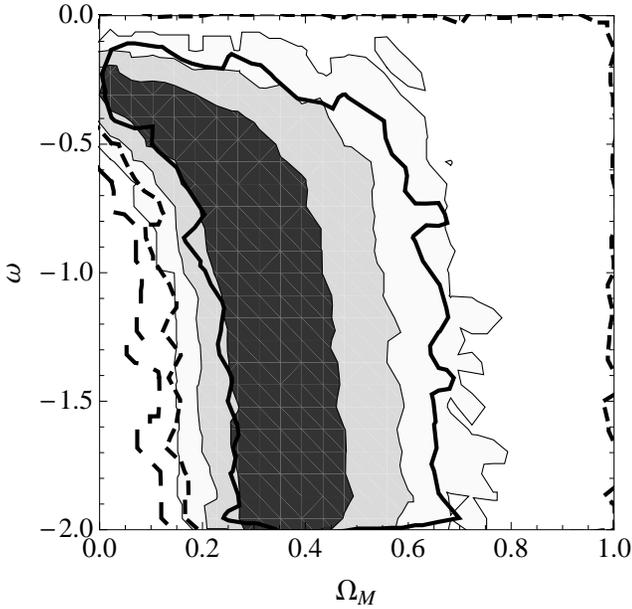}}
 \caption{Marginalised posterior probability for $\Omega_\mathrm{M}$ and $\omega$ as derived from 3 measured time-delays. Grey-shadowed regions are for $\delta D_{\Delta t}^\mathrm{Obs}/D_{\Delta t}^\mathrm{Obs} =10\%$; thick (full, dashed or long-dashed) lines are for $\delta D_{\Delta t}^\mathrm{Obs}/D_{\Delta t}^\mathrm{Obs} =20\%$. Contour values are as in Fig.~\ref{fig_like_statistics_Omega_M_omega}.}
   \label{fig_like_delta_t_Omega_M_omega}
\end{figure}

Time-delay measurements might also constrain dark matter and dark energy. In Fig.~\ref{fig_distances_OmegaM_omega}, $D_{\Delta t}$ is plotted as a function of the cosmological density parameters. Lensing by LISA exploits very distant sources. The dependence on cosmological parameters is then somewhat orthogonal to tests exploring a lower redshift range, such as observations of type Ia supernovae. This can be seen by comparing constant contours of either $D_{\Delta t}$ or luminosity distance in the region of interest of the parameter space. 

Since the total number of lenses is small, lensing methods by their own sample only a restricted redshift interval and are not able to break the degeneracy in the $\Omega_\mathrm{M}$-$\omega$ plane. As can be seen in Fig.~\ref{fig_distances_OmegaM_omega}, even with an extremely accurate but single measurement of $D_{\Delta t}$, the degeneracy on the dark energy equation of state would be severe. A better observational accuracy could only reduce the confidence regions which would still be very elongated along the degeneracy direction. Even with a very small error on $D_{\Delta t}$, the accuracy on $\Omega_\mathrm{M}$ is still affected by the degeneracy. The degeneracy can be broken only by combining with other tests exploring different redshifts.

Together with $\Omega_\mathrm{M}$ and $\omega$ we have to consider as unknown model parameters the redshifts of the lensed sources. For our optimistic lensing scenario, we have then to add 3 unknown lens redshifts to the 2 cosmological parameters. We explored the parameter space by running four Markov chains of $5\times 10^4$ samples each. This was more than enough to reach convergence, with the Gelman and Rubin ratio \citep{ge+ru92,le+br02} being well below 1.01 for each parameter. The prior for each source redshift was uniform and non zero for $z_\mathrm{s} >z_\mathrm{d}$.

The marginalised posterior probability function for $\Omega_\mathrm{M}$ and $\omega$ is plotted in Fig.~\ref{fig_like_delta_t_Omega_M_omega}. After marginalization, the uncertainty on $\Omega_\mathrm{M}$ is $\sim 0.11$ or $\sim 0.21$ for $\delta D_{\Delta t}^\mathrm{Obs} /D_{\Delta t}^\mathrm{Obs} = 10\%$ or $20\%$, respectively. Since the redshift range is the same and the time-delay test is still based on angular diameter distances, the degeneracy in the parameter space is similar to lensing statistics. 

A less optimistic expectation for the accuracy in the determination of the luminosity distances translate into worse cosmological constraints. However, the effect is not so dramatic. Assuming a 20\% relative error on $D_\mathrm{L}$, the uncertainty on $\Omega_\mathrm{M}$ after marginalization would be $\sim 0.12$ or $\sim 0.22$ for $\delta D_{\Delta t}^\mathrm{Obs} /D_{\Delta t}^\mathrm{Obs} = 10\%$ or $20\%$.

The time-delay test might be effective even for a single detected lensing event, when we get $\delta \Omega_\mathrm{M} \sim 0.18$ for $\delta D_{\Delta t}^\mathrm{Obs} /D_{\Delta t}^\mathrm{Obs} = 10\%$.

\subsection{Lensing statistics and time-delays}

\begin{figure}
 \resizebox{\hsize}{!}{\includegraphics{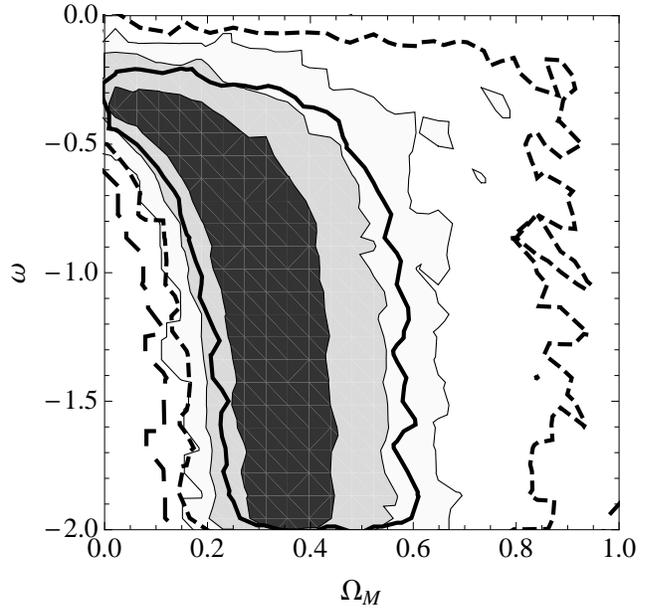}}
 \caption{Marginalised posterior probability for $\Omega_\mathrm{M}$ and $\omega$ as derived from both lensing statistics and time-delays in the case of 3  lensing events. Regions and contours are as in Fig.~\ref{fig_like_delta_t_Omega_M_omega}.}
   \label{fig_like_combined_Omega_M_omega}
\end{figure}

\begin{figure}
 \resizebox{\hsize}{!}{\includegraphics{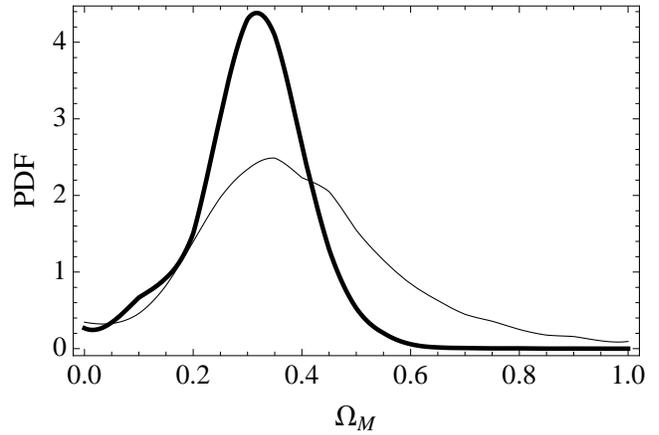}}
 \caption{Marginalised posterior probability for $\Omega_\mathrm{M}$ as derived from both lensing statistics and time-delays in the case of three lensing events. The thick and the thin lines are for $\delta D_{\Delta t}^\mathrm{Obs} /D_{\Delta t}^\mathrm{Obs} = 10\%$ or $20\%$, respectively.}
   \label{fig_pdf_combined_Omega_M}
\end{figure}

To consider lensing statistics and time-delays at the same time, we used a combined likelihood ${\cal L} \propto {\cal L}_\mathrm{stat} \times {\cal L}_{\Delta t}$. The parameter space was explored as in the time-delay case. The marginalised posterior probability function for $\Omega_\mathrm{M}$ and $\omega$ is plotted in Fig.~\ref{fig_like_combined_Omega_M_omega}. We assumed that the Hubble parameter is known. The confidence regions are more confined than for lensing statistics or time delay alone, but the equation of state is still undetermined. The marginalised probability for $\Omega_\mathrm{M}$ is plotted in Fig.~\ref{fig_pdf_combined_Omega_M}. 

The dark matter density parameter might be determined with an accuracy of $\sim0.10$  ($ 0.18$) for 3 lenses and $\delta D_{\Delta t}^\mathrm{Obs} /D_{\Delta t}^\mathrm{Obs} = 10\%$  ($20\%$). Due to the small number of lenses, the final accuracy is mainly determined by the time-delay test.

\section{Forecasting for known source redshifts}
\label{know}

\begin{figure}
 \resizebox{\hsize}{!}{\includegraphics{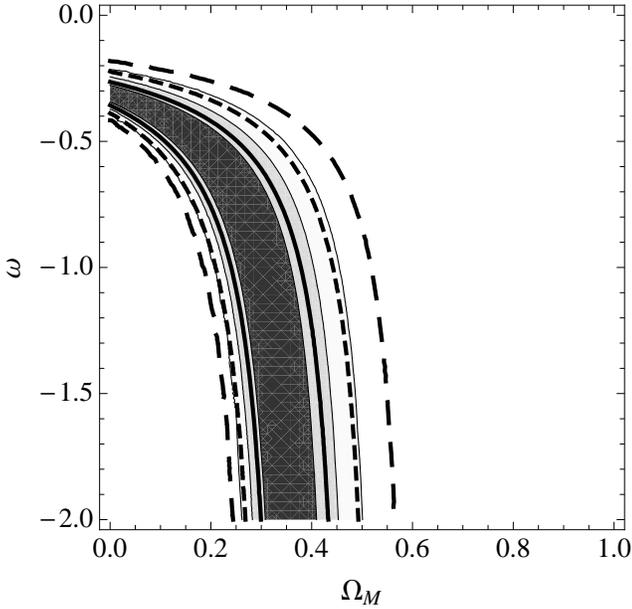}}
 \caption{Posterior probability for $\Omega_\mathrm{M}$ and $\omega$ as derived from 3 measured time-delays of sources with known redshifts. Grey-shadowed regions are for $\delta D_{\Delta t}^\mathrm{Obs}/D_{\Delta t}^\mathrm{Obs} =10\%$; thick lines are for $\delta D_{\Delta t}^\mathrm{Obs}/D_{\Delta t}^\mathrm{Obs} =20\%$. Contour values are as in Fig.~\ref{fig_like_delta_t_Omega_M_omega}.}
   \label{fig_like_delta_t_with_zs_Omega_M_omega}
\end{figure}

\begin{figure}
 \resizebox{\hsize}{!}{\includegraphics{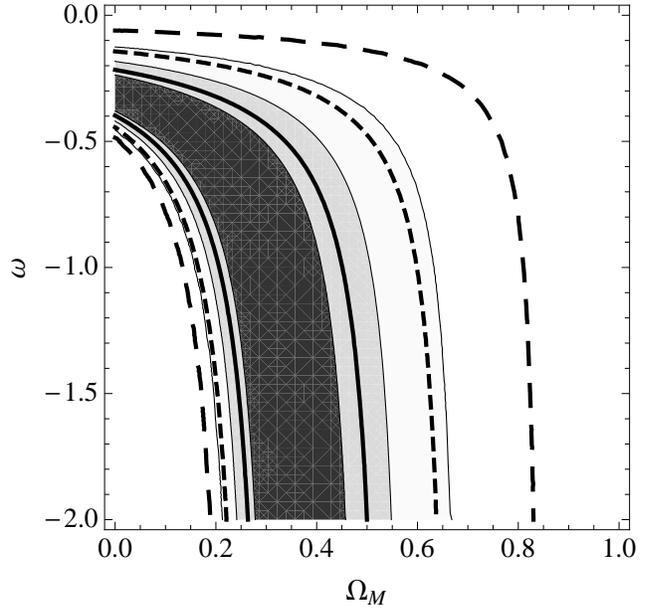}}
 \caption{Posterior probability for $\Omega_\mathrm{M}$ and $\omega$ as derived from 1 measured time-delay of a source with known redshift. Regions and contours are as in Fig.~\ref{fig_like_delta_t_with_zs_Omega_M_omega}.}
   \label{fig_like_delta_t_with_zs_1_lens_Omega_M_omega}
\end{figure}

\begin{figure}
 \resizebox{\hsize}{!}{\includegraphics{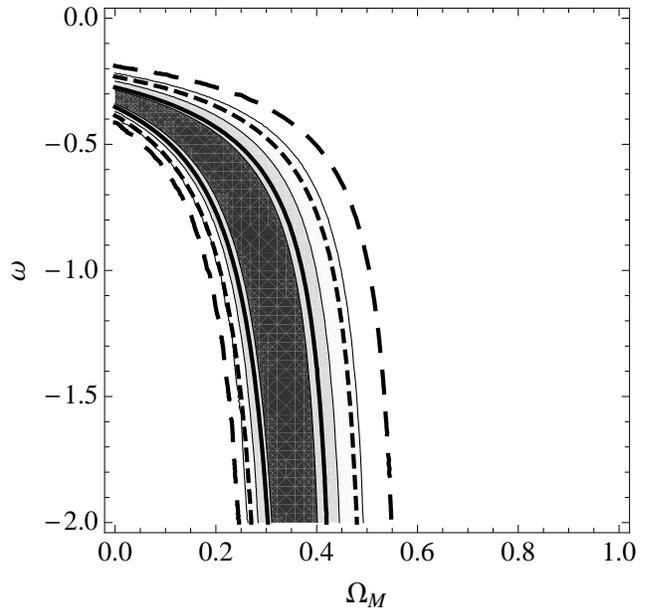}}
 \caption{Posterior probability for $\Omega_\mathrm{M}$ and $\omega$ as derived from both lensing statistics and time-delays in the case of three lensing events with known source redshifts. Regions and contours are as in Fig.~\ref{fig_like_delta_t_with_zs_Omega_M_omega}.}
   \label{fig_like_combined_with_zs_Omega_M_omega}
\end{figure}

In a best-case scenario source redshifts might be determined. In fact, lensing might help in finding the electromagnetic counterparts of the GW signals. Together with the GWs from the coalescence, galaxies harbouring the merging black holes should be lensed as well. Multiple optical images of the host galaxy might be then detected in the nearby of the deflector. The fluxes of the images should be magnified in agreement with the relative amplification of the GW signals, and their position relative to the lens should match the prediction from the analysis of the lensed wave-forms, see Sec.~\ref{dist}. Follow-up observations should make the photometric or spectroscopic determination of the source redshift possible. In principle and very speculatively, the source redshift might be estimated even if only the most magnified optical or radio image of the host galaxy is brought over-threshold. In such a case, the lensed galaxies could not be found as usual by comparing the spectra or the multi-band fluxes of near galaxies, but we should look for a galaxy whose distance from the lens is in agreement with the prediction from the lensing analysis of the lensed GWs.

However, the inclusion of magnification ratios in lensing analyses is problematic for two main reasons \citep{sah+al06}: first, optical flux ratios may be contaminated by microlensing and differential extinction; second, relative magnifications along different directions are weakly coupled with time delays, because magnification measures the local second derivative of the arrival time. Future searches for host galaxies should account for this.

Once the redshift of a lensed source is known, one can strengthen the lensing tests by exploiting the distance-redshift comparison for the source, as usually done when building the Hubble diagram. The additional information on the source redshift makes the lensing constraints on the cosmological parameters much tighter. To estimate the impact of known source redshifts on our forecasting, we re-made the analysis of Sec.~\ref{fore} by fixing the $z_\mathrm{s}$'s to their actual values. Now, the only free parameters to be inferred are the cosmological parameters. In our studio-case, we are then left with just two parameters (either $H_0$  and  $\Omega_\mathrm{M}$ or $\Omega_\mathrm{M}$ and $\omega$). 

Constraints from the analysis of time delays get much stronger. The second term in the right-hand side of Eq.~(\ref{like2}), which before stood to account for an uncertainty and could only help to estimate the source redshift, provides now an additional strong constraint on the cosmological parameters. Posterior probability densities were computed on a two-dimensional grid. 

Uncertainties on $h$ range from 0.1 to 0.3, see Table~\ref{tab_delta_h}. Once the source redshift is fixed, even a single lensing event can provide very interesting bounds. The expected accuracy on $H_0$ from lensed GWs compares very well with more standard results from time-delay analyses. \citet{sah+al06} obtained $\delta h \sim 0.1$ for 10 lensing galaxies.

Results for time-delay constraints on dark matter and dark energy are represented in Fig.~\ref{fig_like_delta_t_with_zs_Omega_M_omega} for the case of 3 lenses and in Fig.~\ref{fig_like_delta_t_with_zs_1_lens_Omega_M_omega} for 1 lens. Thanks to the additional constraint from the distance-redshift relation, confidence regions in the $\Omega_\mathrm{M}$-$\omega$ plane are much more restricted than for unknown source redshifts. Due to the degeneracy, the accuracy improvement for the determination of $\Omega_\mathrm{M}$ is not so significant. With three lenses, the uncertainty on $\Omega_\mathrm{M}$ is $\sim 0.08$ or $\sim 0.09$ for $\delta D_{\Delta t}^\mathrm{Obs} /D_{\Delta t}^\mathrm{Obs} = 10\%$ or $20\%$, respectively. Improvement is more significant for the case of only one lens, when the uncertainty on $\Omega_\mathrm{M}$ is $\sim 0.11$ or $\sim 0.12$ for $\delta D_{\Delta t}^\mathrm{Obs} /D_{\Delta t}^\mathrm{Obs} = 10\%$ or $20\%$, respectively.

The lensing statistics test is not very helpful in such an optimistic scenario. Combined posterior probability densities are plotted in Fig.~\ref{fig_like_combined_with_zs_Omega_M_omega}. Confidence regions are very similar to those obtained exploiting only the time delays.

\section{Conclusions}
\label{conc}

The serendipitous discovery of multiple images of gravitational wave sources by LISA might offer new possibilities for astronomical investigations. Gravitational lensing has been long considered as a tool for cosmography. Here, we considered the peculiarities of cosmological tests with lensed GW sources. Both time-delay measurements and lensing statistics were investigated, which required to develop a treatment of lensing probabilities for transient events. 

A couple of features of the discussed methods deserves particular attention. The main appeal of lensing methods is that they do not need the electromagnetic counterpart to be identified. Classical lensing tests can be developed with knowledge of either the source redshift or the distance. This is different from the usual cosmographic approach proposed for LISA, which attempts to build up the Hubble diagram from the measured distances to the binaries. Inspiral GWs encode the luminosity distance to a binary but they do not encode the source cosmological redshift. To build the Hubble diagram, the electromagnetic counterpart of the GW emission needs to be localised independently to determine the event redshift. Lensing methods can overcome such a shortcoming.

LISA sources are expected to lie at very high redshifts. Even if the accuracy of the proposed lensing tests is not competitive with results from type Ia supernovae or the cosmic microwave background, the explored redshift range would be quite unique and cosmological parameters might be tested with direct methods out to $z \ls 10$. The proposed methods are still based on cosmological distances but, due to the distinct redshift range explored, are orthogonal to other tests at lower $z$. 

This circumstance could make very interesting their use in combination with other techniques. Let us just consider a still very mild constraint on the dark energy equation of state, whose upper allowed limit is lowered from $0$ to $-0.5$ in the following example. In our optimistic scenario (3 lenses, $\delta D_{\Delta t}/D_{\Delta t}=10\%$), the accuracy on the dark matter parameter from the time-delay test would improve from $\delta \Omega_\mathrm{M}\simeq 0.11$ to $0.09$ for unknown source redshifts or from $\delta \Omega_\mathrm{M}\simeq 0.08$ to $0.06$ for known $z_\mathrm{s}$'s.

At the very least, even in a pessimistic scenario with just one lens the cosmological concordance model might be tested in an independent context and the Hubble constant would be directly measured up to very high redshifts. The ability to identify the lens is crucial to the discussed methods. For high redshift sources with a small signal-to noise ratio, the sky location accuracy might be quite poor, $\sim 1 \deg^2$. Several hundreds of sources are expected to be localised in such error-box. However, to claim lensing two above threshold signals have to be observed in the same sky position. By combining the statistical estimated positions for the two sources, the final error-box should be reduced by half. Lensing theory can also contribute to candidate selection. Deflectors should be around redshift one, where we should consider a comoving volume of $\sim 1-3 \times 10^{10}\mathrm{Mpc}^3$. Lenses are also expected to be very massive, which makes them rare. Considering a numerical density of $\gs 10^{-4}\mathrm{Mpc}^{-3}$, we expect just a dozen of lens candidates in the LISA error box. 

In a more optimistic scenario, the lensed signal might be very loud, which would allow a much more precise determination of the source position. In a best case scenario, lensing amplification of the source could be enough to make the electromagnetic counterpart of the source detectable in both images.

The certified accuracy of cosmological lensing methods with LISA can be obtained only with a good knowledge of the galaxy formation history and merging rates. This is the essential ingredient to estimate the number of multiple lensing events expected to be detected by LISA \citep{ser+al10}. Nevertheless, even in pessimistic scenarios cosmological tests based on gravitational lensing of GWs by LISA seem to offer a new and independent way to investigate the properties of dark matter and dark energy and fix the global distance scale. From an alternative point of view, if we are very confident on the estimates of cosmological parameters as obtained from independent tests, lens number counts with LISA might be used to constrain galaxy formation and evolution up to $z \ls 3$.

\setlength{\bibhang}{2.0em}

\end{document}